\documentclass[12pt]{article}

\usepackage{amsmath}
\usepackage{amssymb}
\usepackage{amsthm}
\usepackage{amsfonts}
\usepackage{latexsym}
\usepackage{cite}
\usepackage{graphicx}

\title{Teleportation via multi-qubit channels} 

\author{ 
Jon Links$^{\spadesuit\dagger\sharp}$, John Paul Barjaktarevic$^\S$, \\
Gerard J. Milburn$^{\S\dagger\P}$ and Ross H. McKenzie$^{\S\dagger}$\\
$^\spadesuit$Department of Mathematics, \\
$^\S$Department of Physics, \\
$^\dagger$Centre for Mathematical Physics, \\
$^\P$Centre for Quantum Computer Technology, \\
School of Physical Sciences, The University of Queensland, 4072\\
Australia \\
$^\sharp $email: jrl@maths.uq.edu.au 
}

\begin{document}
\maketitle

\def\L{\mathcal{L}}
\begin{abstract}
We investigate the problem of teleporting an unknown qubit state to a 
recipient via a channel of $2\L$ qubits. In this procedure a protocol is 
employed whereby $\L$ Bell state measurements are made and information based 
on these measurements is sent via a classical channel to the recipient. 
Upon receiving this information the recipient determines a local gate which is 
used to recover the original state. We find that the $2^{2\L}$-dimensional 
Hilbert space of states available for the channel admits a decomposition 
into four subspaces. Every state within a given subspace is a perfect 
channel, and each sequence of Bell measurements projects $2\L$ qubits 
of the system into one of the four subspaces. As a result, only two bits of classical information need be 
sent to the recipient for them to determine the gate. We note some connections
between these four subspaces and ground states of many-body Hamiltonian systems, and discuss the 
implications of these results towards understanding entanglement 
in multi-qubit systems.   \\
~~~\\
~~~\\
%Keywords:  

\end{abstract}
%\pacs{02.30.Ik, 03.75.Hh } 

\def\oR{R^*} \def\upa{\uparrow}
\def\R{\check{R}} \def\doa{\downarrow}
\def\oL{\overline{\Lambda}}
\def\nn{\nonumber} \def\dag{\dagger}
\def\be{\begin{equation}}
\def\ee{\end{equation}}
\def\beq{\begin{equation}}
\def\eeq{\end{equation}}
\def\bea{\begin{eqnarray}}
\def\eea{\end{eqnarray}}
\def\ve{\vec{\epsilon}}
\def\si{\sigma}
\def\th{\theta} \def\ga{\gamma}
\def\l{\left}
\def\r{\right}
\def\a{\overline{a}}
\def\b{\overline{b}}
\def\g{\gamma}
\def\La{\Lambda}
\def\w{w_1,w_2}
\def\u{u_1,u_2}
\def\v{\overline{v}}
\def\n{\vec{n}}
\def\k{\vec{k}}
\def\o{\overline}
\def\rr{\mathcal{R}}
\def\T{\mathcal{T}}
\def\F{\mathcal{F}}
\def\U{\overline{U}}
\def\X{\overline{X}}
\def\N{\mathcal{N}}
\def\p{\partial}
\def\J{\mathcal J}
\def\P{\mathcal{P}}
\def\sop{\mathcal{S}}
\def\top{\mathcal{T}}
\def\Q{\overline{Q}}
\def\RR{\overline{\R}}
\def\ll{\cal l}
\def\]{\}}
\def\vepsilon{\varepsilon}
\def\ve{\varepsilon}
\def\EE{\mathcal{E}}
\def\X{\tilde{X}}
\vfil\eject
\section{Introduction} 

In recent times entanglement has come to be recognised as one of the 
major distinguishing
features between quantum systems and classical
systems, where it is now seen as being as fundamental as the
uncertainty principle. This point of view has arisen due to the realisation 
that
entanglement is a resource to be exploited in the processing
of quantum information \cite{nc} 
through processes such as teleportation \cite{bbcjpw},
dense-coding \cite{bw} and quantum cryptography \cite{k}.
It has also
opened new perspectives in other areas such as condensed matter physics,
due the to emerging understanding of the relationship between entanglement and quantum critical
phenomena \cite{on,oaff,vidal,sobkv}. 
As a consequence there has been an intense level
of activity in characterising entanglement and studying its properties. 

At the level of bi-partite
systems entanglement is well understood and can be quantified 
\cite{bbps,pr,hw,w,n}. 
{}From studies of three-qubit systems it was
realised \cite{ckw,dvc} that different categories of entanglement exist in
multi-qubit systems,  
with the specific example of three-way entanglement shown to be essentially
different from bi-partite entanglement through the examples of the
Greenberger-Horne-Zeilinger (GHZ) and W states. 
Now, a clear picture of
three-qubit entanglement has emerged with the demonstration of three different 
types of entanglement existing in the three-qubit case, which are
characterised by five generically independent invariants \cite{s,g}. 
Though the above results for three-qubit systems 
can in principle be generalised to
arbitrary multi-qubit systems,  
it is technically challenging to undertake. Despite many
studies of specific types of multi-qubit entanglement 
(e.g. \cite{clps,sobkv,ckw,vw,wz,wc,mw,jsst,lt,heb,bssb}), 
a complete description remains elusive. 

\begin{figure}[h]
\begin{center}
\includegraphics[width=0.45\textwidth]{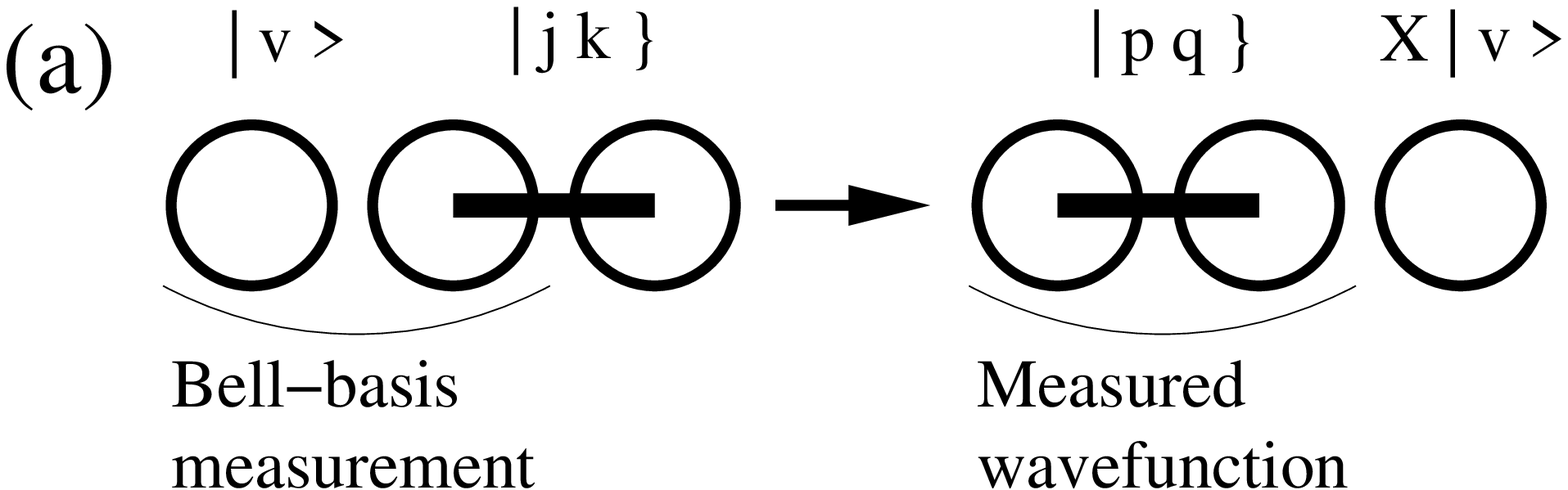} \\
~~\\
\includegraphics[width=0.7\textwidth]{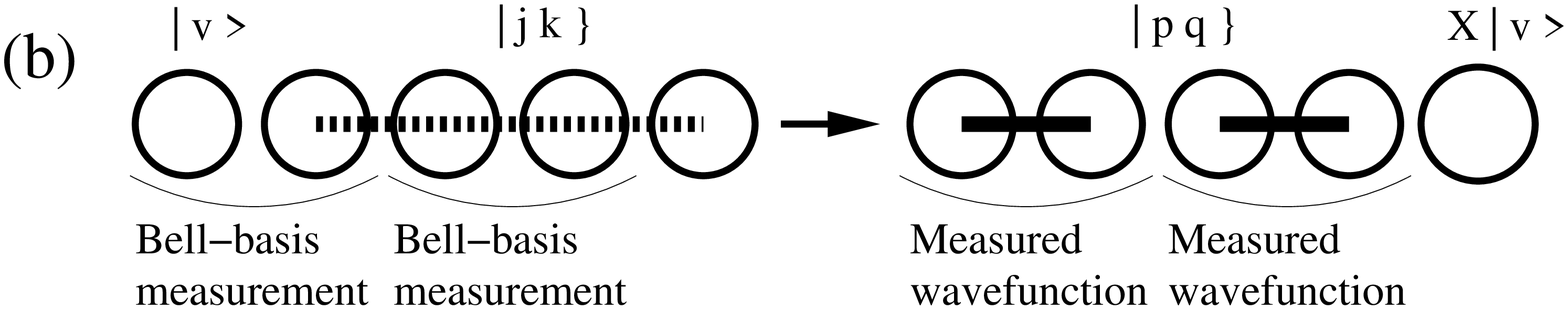}
\end{center}
\caption{Schematic representation of the teleportation protocol. Circles denote
qubit states and two circles 
joined by a solid line denotes a Bell state. Multi-qubit perfect
channels are represented by circles joined by a dashed line. (a) 
For the case of a 2-qubit channel with client state $\l|v\r>$, 
a Bell measurement is made on the
subsystem comprised of the client qubit and the 
first qubit of the channel. After 
measurement, the third qubit is left in the state 
$X\l|v\r>$, where $X$ is a unitary operator 
dependent on the measurement outcome. 
(b) In the case of a 4-qubit channel, two Bell measurements
leave the final qubit in the state $X\l|v\r>$, 
where $X$ is determined by both measurement outcomes.
In general for a $2\L$-qubit channel, a sequence of $\L$
Bell measurements needs to be made to implement the protocol.}
\label{fig0}
\end{figure}

Our aim in this work is to investigate entanglement in
multi-qubit systems through a study of one of its applications, viz.
teleportation.  The protocol
for the procedure is as follows, with a schematic representation shown in Fig. (\ref{fig0}). 
An unknown qubit state is held by a 
client (Alice), 
and is to be teleported to a recipient (Bob). Alice and 
Bob share a quantum channel which is some state of $2\L$ qubits, 
so the entire system consists of $2\L+1$
qubits. The channel is distributed in such a way that Alice may access
$2\L-1$ qubits of the channel while Bob has access to a single
qubit of the channel. Alice is to perform $\L$ Bell state measurements on the
$2\L$-qubit subsystem which is comprised of her unknown state and $2\L-1$
qubits of the channel. The consequence of this measurement is that 
Bob is left with a single qubit which is not entangled with the
remainder of the system. From the results of the measurements
 Alice is to send classical information to
Bob. Upon receiving this information, and some
knowledge of the channel, Bob determines a
local unitary operation called a {\it correction gate} which he applies to
his qubit. Any channel for which this procedure exactly reproduces the
client state for Bob (i.e. the teleportation is effected with
perfect fidelity) we will call a {\it perfect channel}.  One of the aims
of this work is to determine the complete set of perfect channels for this protocol.
We mention that this protocol is not {\it tight} in the sense of 
\cite{werner}, and consequently does not belong to   
the classification of teleportation schemes given therein. 
On the other hand it does bear similarity to the {\it quantum repeater} described in \cite{pvmc}, 
with the major difference being that we employ Bell measurements whereas local measurements are used in \cite{pvmc}.

It is well known that teleportation can be performed with perfect fidelity
across a 2-qubit channel when the channel is one of the four Bell states
\cite{bbcjpw}. This is achieved by making a Bell state measurement and
then sending two bits of information to the recipient via a classical
channel, which is then used to determine the correction gate. It is thus
clear that teleportation can also be achieved with perfect fidelity using
a channel which is a product of $\L$ Bell states, by performing $\L$
successive Bell state measurements. However this is not
the most general solution to the problem we have described above, and 
our analysis below shows some surprising
results. The first is that there exist four orthogonal perfect channel
subspaces, the direct sum of which is the entire Hilbert space of
channels. Also, despite the fact that $\L$ Bell measurements need to
be performed by Alice to implement the teleportation, 
only two bits of classical information need
to be sent from Alice to Bob for him to determine the correction gate.

Because perfect channels fall into one of only four subspaces of the
channel state space, an interesting question to consider is whether the ground states of
common many-body systems fall into these classes. This is indeed the case.
For example, our results indicate that all spin singlet states are perfect
channels, and so the ground state of the antiferromagnetic Heisenberg
model, for a number of different lattices, is a perfect channel, 
as is the ground state of the one-dimensional Majumdar--Ghosh model \cite{mg}. It is also true 
that the ground state of the model of Affleck, Kennedy, Lieb and
Tasaki (AKLT) \cite{aklt1,aklt2} is a perfect channel, under an equivalent protocol 
\cite{pvmc,vmc,fkr}. 
In identifying the perfect channel states we determine a
teleportation-order parameter which provides a measure of the
effectiveness of an arbitrary channel. The teleportation-order parameter 
has close connection with string-order \cite{dnr}, as discussed in 
\cite{pvmc,vmc} in relation to the AKLT model, and is also an example of the string operators
discussed in \cite{kitaev}. We mention however that the results of our analysis are independent
of the dimension and topology of the lattice on which the qubits are arranged.   
 
We will
also show that this analysis extends to formulate a teleportation protocol
for the case of 3-qubit channels, and that there is a generalisation for qudits.  
Finally, we will discuss some 
implications of
these results towards understanding entanglement 
in multi-qubit systems. The results presented here describe 
in detail the mathematical 
aspects which underly the 
results reported in \cite{bmlm}.

\section{Teleportation via two-qubit channels}  

In this section we recall teleportation across a channel of two
qubits which exists in one of the four Bell states \cite{bbcjpw}. While this phenomenon 
is now well known, 
the notational conventions we adopt, which are
convenient for the following sections, are not
standard. 
 
Let $\l|+\r>,\,\l|-\r>$ denote the standard
basis for a qubit space $V$ such that $\l|j\r>$ is an eigenvector 
of the Pauli matrix
$\sigma^z$ with eigenvalue $j$. Throughout, we will label $\pm 1$
simply by $\pm$.  A natural basis for two coupled qubits is 
$\l|j,k\r>\equiv \l|j\r>\otimes\l|k\r>$. 
In making a basis change to the Bell states we define  
\bea 
\l|+:+\r\}&=&\frac{1}{\sqrt{2}}\left(\left|+,+\right>
+\left|-,-\right>\right) \nn
\\
\l|+:-\r\}&=&\frac{1}{\sqrt{2}}\left(\left|+,-\right>
+\left|-,+\right>\right) \nn \\
\l|-:+\r\}&=&\frac{1}{\sqrt{2}}\left(\left|+,+\right>
-\left|-,-\right>\right) \nn \\
\l|-:-\r\}&=&\frac{1}{\sqrt{2}}\left(\left|+,-\right>
-\left|-,+\right>\right) \nn \eea  
such that we can write  
\bea 
\l|j:k\r\}&=&\frac{1}{\sqrt{2}}\l(\l|+,k\r>+j\l|-,\o{k}\r>\r)
\label{compact}  \eea 
where we adopt the notation $\o{k}=-k$.
It is known that each of the Bell basis states are related by a
{\em local} unitary transformation, which we express as 
\bea \l|j:k\r\}=(I \otimes X^{jk}_{pq}) \l|p:q\r\} \nn \eea  
where
\bea
&& X^{jk}_{jk}=U^0 \nn \\
&& X^{+-}_{++}= X^{++}_{+-} = X^{-+}_{--}=X^{--}_{-+} = U^1 \nn \\
&& X^{-+}_{++}= X^{++}_{-+}=  -X^{+-}_{--}=-X^{--}_{+-}= U^2 \nn \\
&& X^{--}_{++}= X^{+-}_{-+}= -X^{++}_{--}=-X^{-+}_{+-}= U^3,  \label{x=u} \eea
and the unitary operators $U^i$ are given by  
\bea
U^0&=&\begin{pmatrix}1& 0 \cr  0 & 1\end{pmatrix} \nn \\
U^1&=&\begin{pmatrix}0& 1 \cr  1 & 0\end{pmatrix} \nn \\
U^2&=&\begin{pmatrix}1& 0 \cr  0 & -1\end{pmatrix} \nn \\
U^3&=&\begin{pmatrix}0& -1 \cr  1 & 0\end{pmatrix} \nn \eea
(Note that the $X^{jk}_{pq}$ could just as easily have been defined in 
terms of Pauli matrices. We generally prefer to not use Pauli matrix notation, 
as this eliminates $\sqrt{-1}$ terms which would otherwise appear 
in many subsequent formulae.) The above is just a 
statement of the fact that the Bell states are {\it equivalent}: two states 
are said to be equivalent if they are equal up to a tensor product of 
local unitary transformations. Equivalent states have identical entanglement properties. 
Likewise, we say that two subspaces $Y,\,Z$ with the same (finite) dimension are equivalent if and 
only if for a fixed transformation, each $y\in\,Y$ is equivalent to some $z\in\,Z$. Moreover two operators are 
equivalent if they are similar by a transformation which is a tensor 
product of local unitary transformations.  

Define $\nu: \left\{ \pm
1\right\} \rightarrow \mathbb{Z}_2$ by
\bea \nu(+1)=0, && ~~~~~~\nu(-1)=1 \nn \eea
which satisfies $\nu(ab)=\nu(a)+\nu(b)$.
We can express the relations (\ref{x=u}) as 
\bea X^{jk}_{pq} &=& \delta^{jk}_{pq}(U^1)^{\nu(kq)}(U^2)^{\nu(jp)}  
\label{factor} 
\eea 
where $\delta^{jk}_{pq}=\pm 1$ can be read off from (\ref{x=u}). 
The operators $X^{jk}_{pq}$ satisfy the following properties:
\bea
\phantom{X^{jk}_{pq} }X^{jk}_{pq}&=& \ve^{jk}_{pq} X^{pq}_{jk}
=(X^{pq}_{jk})^\dagger \label{p1} \\
X^{jk}_{pq} X^{pq}_{ab} &=& X^{jk}_{ab} \label{p2} \\
X^{jk}_{pq} X^{ab}_{cd} &=& \ve^{jk}_{pq} \ve_{jb}^{pq}
X^{jb}_{pq} X^{ak}_{cd} \label{p3} \\
X^{jk}_{pq} X^{ab}_{cd} &=& \delta^{ak}_{jk}\delta^{ab}_{jb}\ve^{jk}_{pq} \ve_{ak}^{pq} 
X^{ak}_{pq} X^{jb}_{cd} \label{p4} \\  
X^{jk}_{pq} X^{ab}_{cd} &=& \delta^{jk}_{pq}\delta^{ab}_{cd}
\delta^{(ja)(kb)}_{(pc)(qd)} 
\ve^{jb}_{pd} X^{(ja)(kb)}_{(pc)(qd)} \label{p5} \eea
where $\ve^{jk}_{pq}=\ve_{jk}^{pq}$ is defined by 
\begin{equation}
\ve^{jk}_{pq}=\left\{\begin{array}{ll}
  -1 & \hbox{ for } j\neq p \hbox{ and } k\neq q\\ 
  \phantom{-} 1  & \hbox{ otherwise } 
      \end{array}\right. 
      \end{equation}
Property (\ref{p1}) is deduced by inspection, while (\ref{p2}) follows
from the definition of the $X^{jk}_{pq}$. To show
(\ref{p3}),
first observe that it is true if $k=b$. Assuming $k\neq b$ and using
(\ref{p1},\ref{p2}) we find
\bea
X^{jk}_{pq} X^{ab}_{cd} &=& \ve^{jk}_{pq} 
X^{pq}_{jk} X^{ab}_{cd} \nn \\
&=& \ve^{jk}_{pq} X^{pq}_{jb} X^{jb}_{jk} X^{ab}_{ak} X^{ak}_{cd} \nn \\
&=& \ve^{jk}_{pq} X^{pq}_{jb} U^1 U^1 X^{ak}_{cd} \nn \\
&=& \ve^{jk}_{pq} \ve_{jb}^{pq} X^{jb}_{pq} X^{ak}_{cd}.  \nn \eea
Property (\ref{p4}) is proved similarly.
To show (\ref{p5}) we calculate 
\bea X^{jk}_{pq} X^{ab}_{cd} &=&\delta^{jk}_{pq}\delta^{ab}_{cd} 
(U^1)^{\nu(kq)} (U^2)^{\nu(jp)}
(U^1)^{\nu(bd)} (U^2)^{\nu(ac)} \nn \\ 
&=& \delta^{jk}_{pq}\delta^{ab}_{cd} \ve^{jb}_{pd}
(U^1)^{\nu(kqbd)}(U^2)^{\nu(jpac)} \nn \\ 
&=& \delta^{jk}_{pq}\delta^{ab}_{cd} \delta^{(ja)(kb)}_{(pc)(qd)} 
\ve^{jb}_{pd}  X^{(ja)(kb)}_{(pc)(qd)}.   \nn \eea   
We also find that
\bea
U^1\otimes U^1 \l|j:k\r\}&=&j  \l|j:k\r\} \nn \\
U^2\otimes U^2 \l|j:k\r\}&=& k \l|j:k\r\} \nn \eea
so the eigenvalues of $U^1\otimes U^1,\,U^2\otimes U^2$   
provide good quantum numbers to label the basis states. 
(It is easily checked that
$U^1\otimes U^1$ and $U^2\otimes U^2$ commute.) 
A measurement which is represented by the 
action of these two operators is called a Bell measurement, 
where $(j:k)$ denotes the measurement outcome.

Let $\left|v\r>=\alpha\l|+\r>+\beta\l|-\r>$ be arbitrary such that
$$|\alpha|^2+|\beta|^2=1$$  and $\alpha$ and $\beta$ are completely
unknown.  We call $\l|v\r>$  the {\em client state}.
The state which will be used to teleport the client state will be called
the {\em channel}. When the channel is one of the Bell basis
states $\l|j:k\r\}$ we look to rewrite the total state $\l| v\r>\otimes \l|j:k\r\}$ 
as a linear combination of states where the first two qubits 
are expressed in the Bell basis, i.e.    
\bea
&&{2}(I \otimes I \otimes (X^{jk}_{++})^{-1})
\l|v\r> \otimes \l|j:k\r\} \nn \\
&&~~=  {2}\l|v\r> \otimes \l|+:+\r\} \nn \\
&&~~= \sqrt{2}\l(\alpha\l|+,+,+\r>+\beta\l|-,+,+\r> + \alpha\l|+,-,-\r>
+\beta\l|-,-,-\r>
\r)\nn \\
&&~~=\l|+:+\r\}\otimes (\alpha\l|+\r>+\beta\l|-\r>)   
+\l|+:-\r\}\otimes (\beta\l|+\r>+\alpha\l|-\r>) \nn \\
&&~~~~~~~~~~+\l|-:+\r\}\otimes (\alpha\l|+\r>-\beta\l|-\r>) 
+\l|-:-\r\}\otimes (-\beta\l|+\r>+\alpha\l|-\r>)   \nn \\
&&~~=\l|+:+\r\}\otimes  \l|v\r>   
+\l|+:-\r\}\otimes U^1 \l|v\r> %\nn \\
%&&~~~~~~~~~~
+\l|-:+\r\}\otimes U^2 \l|v\r> 
+\l|-:-\r\}\otimes U^3 \l|v\r>    \nn \\
&&~~=\l|+:+\r\}\otimes X^{++}_{++} \l|v\r>   
+\l|+:-\r\}\otimes X_{++}^{+-} \l|v\r> \nn \\
&&~~~~~~~~~~+\l|-:+\r\}\otimes X_{++}^{-+} \l|v\r> 
+\l|-:-\r\}\otimes X_{++}^{--} \l|v\r>    \nn \\
&&~~=\l|+:+\r\}\otimes X^{++}_{++} \l|v\r>   
+\l|+:-\r\}\otimes X^{++}_{+-} \l|v\r> \nn \\
&&~~~~~~~~~~+\l|-:+\r\}\otimes X^{++}_{-+} \l|v\r> 
+\l|-:-\r\}\otimes \ve_{++}^{--}X^{++}_{--} \l|v\r>    \nn
\eea
and so
\bea 2\l|v\r>\otimes \l|j:k\r\}
&=&\l|+:+\r\}\otimes X^{jk}_{++} \l|v\r>   
+\l|+:-\r\}\otimes X^{jk}_{+-} \l|v\r> \nn \\
&&~~~~~~+\l|-:+\r\}\otimes X^{jk}_{-+} \l|v\r> 
+\l|-:-\r\}\otimes \ve_{++}^{--} X^{jk}_{--} \l|v\r>    \nn\\
&=&\l|+:+\r\}\otimes \ve_{++}^{++}X^{jk}_{++} \l|v\r>   
+\l|+:-\r\}\otimes \ve_{++}^{+-}X^{jk}_{+-} \l|v\r> \nn \\
&&~~~~~~+\l|-:+\r\}\otimes \ve_{++}^{-+}X^{jk}_{-+} \l|v\r> 
+\l|-:-\r\}\otimes \ve_{++}^{--}X^{jk}_{--} \l|v\r>.    \nn
\eea
This last expression can be expressed in a compact form: 
\bea  \l|v\r> \otimes \l|j:k\r\}= \frac{1}{2}\sum_{p,q} \l|p:q\r\}\otimes
\X^{jk}_{pq}\l|v\r> \label{comm} \eea 
where $\X^{jk}_{pq}= \ve_{++}^{pq} X^{jk}_{pq}$. Thus, when a Bell measurement is made
on the first and second qubits by Alice, the system is projected onto a state 
$$\l|p:q\r\}\otimes \X^{jk}_{pq}\l|v\r>.$$ 
Note that the probabilities for measuring each of the four possible 
states are equal. The result of the measurement may be communicated to 
Bob using only two bits of classical information. This, together with 
knowledge of which channel was used, 
is sufficient information for Bob to determine
the correction gate $\X^{pq}_{jk}$, to be implemented  
in order to recover the client state. Thus the client state has been
teleported from Alice to Bob via the channel and classical communication.

\section{Teleportation via multi-qubit channels}
\subsection{Singlet channels: an example of teleportation via  
multi-qubit channels with perfect fidelity}   

Our goal is to extend the above construction to the multi-qubit channel 
case. Here, we will first look at the case when the channel is a 
$U(2)$ singlet. The Hilbert space for an $L$-qubit system is 
given by the tensor product of the local qubit spaces $V$; 
$$V^L\equiv V^{\otimes L}=V_1\otimes V_2\otimes \dots \otimes V_L. $$ 
Throughout we take $L$ to be even and define $\L=L/2$.  Recall 
that the action of the Lie group 
$U(2)$ on the space of a single qubit space $V$ is represented 
by the set of all $2\times 2$ unitary matrices. 
Given any such matrix $A\in U(2)$, the action extends to the 
space of $L$ qubits through 
$$A\rightarrow A^{\otimes L}. $$ A $U(2)$ singlet is any state 
$ \l|\Psi\r> \in V^L$ such that for all $A\in \,U(2)$ 
\bea A^{\otimes L} \l|\Psi\r> = \exp(i\theta) \l|\Psi\r> \label{singletaction} \eea 
for some real $\theta$. An example of a singlet is given by 
the Bell state $\l|-:-\r\}$.
 
Let $P^{pq}$ denote the projection onto the Bell state
$\l|p,q\r\}$. Each projection can be related to the projection 
onto the $U(2)$ singlet state $\l|-:-\r\}$ through 
\bea P^{pq}=(I\otimes X^{pq}_{--})P^{--}(I\otimes X^{--}_{pq}). \nn \eea     
Now since $\l|-:-\r\}$ is a $U(2)$ singlet then
$P^{--}$ is an invariant operator in the sense that
\bea (A\otimes A) P^{--} (A^{-1}\otimes A^{-1}) = P^{--}  	
~~~~\forall\, A\in U(2), \nn \eea 
that is, the action of  $U(2)$ commutes with $P^{--}$.
It follows that $P^{--}\otimes I^{\otimes {L-2}}$ is an 
invariant operator on the $L$-fold space $V^L$.
An important result we will use subsequently is  
Schur's lemma, which asserts that any
invariant operator maps an irreducible $U(2)$ invariant 
space to an isomorphic space by a scalar multiple \cite{h}.
 
Let $P^{pq}_r$ be the projector $P^{pq}$ 
acting on the $r$th and $(r+1)$ qubits of the
tensor product space and let $(X^{jk}_{pq})_r$ be 
$X^{jk}_{pq}$ acting on the $r$th space.
Let $\left|v \right>$ again be an arbitrary client state, and let the channel 
$\l|\Psi\r>\in\, V^L$ be an arbitrary singlet state. 
We denote the space to which the client state belongs by 
$V_0$. The initial state of the total system is thus 
$$|v^{(0)}\rangle=\left|v\right> \otimes \left|\Psi\right>.$$
Now we employ Schur's lemma, which 
in particular means that 
$$P_0^{--}\left|v\right> \otimes \left|\Psi\right>
=\chi\l|-:-\r\}\otimes |v^{(1)}\rangle $$  
for some scalar $\chi$, 
where $|v^{(1)}\rangle$ is some state in $W^{(1)}=V_2\otimes
V_3\otimes
...V_L$ which is isomorphic to $\left|v\right>$. 
In other words,  
if we decompose $W^{(1)}$ into $U(2)$ spaces then
$|v^{(1)}\rangle$ belongs to a doublet.

Starting with $|v^{(0)}\rangle$, a Bell
measurement is made on $V_0\otimes V_1$, which is denoted by 
the projection $P_0^{pq}$ where 
$(p:q)$ is the result of the measurement. With reference 
to the above
discussions and notational conventions this
means we may write
\bea
P_0^{pq}|v^{(0)}\rangle 
&=&P_0^{pq} \left|v\right>\otimes \left|\Psi\right>\nn \\
&=&(X^{pq}_{--})_1 P_0^{--}  
(X_{pq}^{--})_1\left|v\right>\otimes \left|\Psi\right>
\nn \\
&=&\left(\prod_{r=1}^L (X^{pq}_{--})_r\right) P_0^{--}  
\left(\prod_{r=1}^L (X_{pq}^{--})_r\right)\left|v\right>\otimes \left|\Psi
\right>\nn \\
&=&\left(\prod_{r=1}^L (X^{pq}_{--})_r\right) P_0^{--}  
\left|v\right>\otimes 
\left(\prod_{r=1}^L (X_{pq}^{--})_r\right) \left|\Psi
\right>\nn \\
&=&e^{i\theta}  \left(\prod_{r=1}^L (X^{pq}_{--})_r\right) P_0^{--} 
\left|v\right>\otimes \left|\Psi \right>~~~~({\rm since}~\l|\Psi\r>~{\rm is ~a~ singlet})\nn \\
&=&e^{i\theta}\chi  
\left(\prod_{r=1}^L (X^{pq}_{--})_r\right)\l|-:-\r\}
\otimes|v^{(1)}\rangle~~~({\rm by~Schur's~lemma})\nn
\\
&=&e^{i\theta}\chi (X^{pq}_{--})_1\l|-:-\r\} \otimes
\left(\prod_{r=2}^L (X^{pq}_{--})_r\right)|v^{(1)}\rangle \nn \\
&=&e^{i\theta}\chi\l|p:q\r\}\otimes\left(\prod_{r=2}^L (X^{pq}_{--})_r\right)
|v^{(1)}\rangle. \nn \eea

This procedure can be iterated by taking $l$ consecutive Bell 
measurements to give
$$ P_0^{p_1q_1}P_2^{p_2q_2}....P_{2l-2}^{p_lq_l}
|v^{(0)}\rangle
=\gamma \l|p_1:q_1\r\}\otimes \l|p_2:q_2\r\}\otimes ... \l|p_l:q_l\r\} 
\otimes \left(\prod_{t=l}^{1}\left(\prod_{r=2l}^{L}
(X^{p_t q_t}_{--})_r\right)\right)|v^{(l)}\rangle
$$ 
where $\gamma$ is a scalar. 
In particular
$$ P_0^{p_1q_1}P_2^{p_2q_2}....P_{L-2}^{p_\L q_\L}
|v^{(0)}\rangle
=\gamma \l|p_1:q_1\r\}\otimes \l|p_2:q_2\r\}\otimes ... \l|p_\L :q_\L\r\} 
\otimes \left(\prod_{t=\L}^{1}
(X^{p_t q_t}_{--})_L\right)|v^{(\L)}\rangle.
$$
Note the notation employed means for any
operator $\Xi$
\bea\prod^1_{j=k} \Xi_j = \Xi_k ....\Xi_2\Xi_1. \nn \eea 
In each case $
|v^{(l)}\rangle \in W^{(l)}=V_{2l}\otimes....\otimes V_L$
is isomorphic to $\left|v\right>$ due to Schur's lemma.
However, since $|v^{(\L)}\rangle\in W^{(\L)}=V_L$ and $V_L$ is an irreducible $U(2)$ space, 
we must have
\bea |v^{(\L)}\rangle=\left|v\right>_L. \nn \eea
After the $\L$ Bell basis measurements are made by Alice, Bob needs to apply the
correction gate
\bea D=\prod_{t=1}^\L X^{--}_{p_t q_t} \nn \eea  
to the $L$th qubit in order to recover the client state. In view of (\ref{p5}) we see that 
\bea D=\kappa X^{jk}_{pq} \label{dg} \eea  
where $j=k=(-1)^\L$, $p=\prod_{t=1}^\L p_t$ and $q=\prod_{t=1}^\L q_t$. 
Note that $\kappa=\pm 1$ in
(\ref{dg}) is a function of all indices $p_t$ and $q_t$ and can in 
principle be determined through (\ref{p5}). However its value is 
inconsequential, as it will only alter the corrected state by a 
phase. Throughout, whenever such a phase arises we will generically 
denote it by $\kappa$. For ease of notation, we will not explicitly 
state its dependence on particular indices, although this should be 
clear from the context.  

After Alice has performed the Bell measurements, she need only send 
two bits of classical information, viz. $p$ and $q$, to Bob in order 
for him to determine the correction gate.  
This is a case where  teleportation occurs with perfect
fidelity, and shows that {\it all} singlet states are perfect channels.  
Next we look to extend this result to cover the most general possibilities.

\subsection{A basis of perfect multi-qubit channels}

Our first step to classifying the perfect channels is to 
establish that there exists a basis 
for the $L$-qubit Hilbert space $V^L$ in which each basis state is a perfect channel. 
Since the Bell states provide a basis for $V\otimes V$ it immediately follows that 
the set of all vectors of the form 
\bea  \l|\r.\vec{j}:\vec{k}\l.\r\}=\l|j_1:
k_1\r\}\otimes \dots \otimes \l|j_\L: k_\L\r\} \label{basis} \eea 
forms a basis for $V^L$. 
Through repeated use of (\ref{comm}) we arrive at 
\bea
\l|v\r>\otimes \l|\r.\vec{j}: \vec{k} \l.\r\}
=\frac{1}{2^{\L}}\sum_{\vec{p},\vec{q}}
\l|\r.\vec{p}:\vec{q}\l.\r\}\otimes \X^{j_\L k_L}_{p_\L q_\L}\dots
\X^{j_1k_1}_{p_1q_1} \l|v\r> \label{xbasis} \eea 
where the sum is taken over all
possible values of $\vec{p}$ and $\vec{q}$. By Alice making pairwise
Bell measurements on the first $L$ spaces, a projection is made to a state
\bea
\l|\r.\vec{p}:\vec{q}\l.\r\}\otimes \X^{j_\L k_\L}_{p_\L q_\L}\dots
\X^{j_1k_1}_{p_1q_1} \l|v\r> .
\label{ms} \eea 
Given a basis vector $\l|\r.\vec{j}:\vec{k}\l.\r\}$ we say that it
belongs to the {\it Bell class} $[j:k], \, j,\,k=\pm $ if 
\bea
\prod_{i=1}^\L j_i = j, ~~~~
\prod_{i=1}^\L k_i = k . \nn \eea 
There are four distinct Bell classes. 
Given an
arbitrary vector 
\bea   \l|\Phi\r>=\sum_{\vec{j},\vec{k}}
\Gamma_{\vec{j},\vec{k}} \l|\r.\vec{j}: \vec{k}\l.\r\} \label{lc}
\eea  
we say that $\l|\Phi\r>$  belongs to the Bell class $[j:k]$ if, for  
$\Gamma_{\vec{j}\vec{k}}$ non-zero, then
$\l|\r.\vec{j}:\vec{k}\l.\r\}\in\,[j:k]$. 
In other words,
$\l|\Phi\r>$ belongs to the Bell class $[j:k]$ if it is a linear combination of
basis vectors (\ref{basis}) of Bell class $[j:k]$. 
It is clear that the notion of Bell classes leads to a vector
space decomposition
$$V^L= V_{[+:+]}^L\oplus V_{[+:-]}^L\oplus V_{[-:+]}^L \oplus V_{[-:-]}^L
$$
and we refer to each $V^L_{[j:k]}$ as a Bell subspace.  
In view of
(\ref{xbasis}) we arrive at
\bea  \l|v\r>\otimes \l|\Phi \r>
=\frac{1}{2^{\L}}\sum_{\vec{p},\vec{q},\vec{j},\vec{k}}
\Gamma_{\vec{j}\vec{k}}\l|\r.\vec{p}:\vec{q}\l.\r\}\otimes \X^{j_\L
k_L}_{p_\L q_\L}\dots \X^{j_1k_1}_{p_1q_1} \l|v\r> \label{needed} \eea 
where the
sum is taken over all possible values of $\vec{j},\,\vec{k},\,
\vec{p}$ and $\vec{q}$. Making pairwise measurements on the first
$L$ spaces then projects out a state 
\bea 
\N \sum_{\vec{j},\vec{k}}\Gamma_{\vec{j}\vec{k}}\l|\r.\vec{p}:\vec{q}\l.\r\}
\otimes \X^{j_\L k_\L}_{p_\L q_\L}\dots \X^{j_1k_1}_{p_1q_1} \l|v\r> 
\label{projected} \eea 
where $\N$ is a normalisation factor. 
Now suppose $\l|\Phi\r> \in \, V_{[j:k]}^L$.  Again appealing to (\ref{p5})
and using the fact that 
$\Gamma_{\vec{j}\vec{k}}=0$ for $\l|\r.\vec{j}:\vec{k}\l.\r\}\notin\,[j:k]$, 
then up to a phase we can express (\ref{projected}) as 
\bea
\l|\r.\vec{p}:\vec{q}\l.\r\}\otimes 
\X^{jk}_{pq} \l|v\r> 
\nn \eea 
where $[j:k]$ is the Bell class of the channel, and $[p:q]$ is the Bell class of the measurement (more precisely, the Bell class of the measured tensor product of Bell states). As in the case of 
singlet channels, Alice again just needs to communicate the Bell class of her measurement 
(i.e. two bits of classical information) to Bob for him to determine the 
correction gate. Here we assume, as in the case of teleportation across a single Bell state, 
that the Bell class of the channel is known to Bob.  

A characteristic of the states
of Bell subspaces is that they are simultaneous 
eigenvectors of the operators 
\bea 
\Upsilon^1&=& \prod_{p=1}^L  U^1_p, 
~~~~~\Upsilon^2= \prod_{q=1}^L  U^2_q,  \nn  \eea  
and therefore an eigenstate of the product 
\bea 
\Upsilon^3 &=& \l(\prod_{p=1}^L  U^1_p\r)\l(\prod_{q=1}^L  U^2_q\r)  
= \prod_{p=1}^L U^3_p.  \nn \eea   
Note also that 
\bea [\Upsilon^\alpha,\, \Upsilon^\beta]=0 ~~~~~~\forall\,\alpha,\,\beta=1,2,3. \nn \eea 
It is apparent that each space $V^L_{[j:k]}$ is a stabiliser space for the 
set of operators $\{j\Upsilon^1,\,k\Upsilon^2\}$. In fact our protocol can 
be re-expressed as a multi-qubit generalisation of the stabiliser description 
of teleportation given in \cite{gott}.
Another result that can be immediately deduced from the above is that 
any perfect channel $\l|\varphi\r>\,\in V^L_{[j:k]}$ is maximally locally disordered; i.e.,
\bea 
\l<\varphi\r|U^\alpha_p\l|\varphi\r>=0~~~\forall \,\alpha=1,2,3,~~\forall\,p=1,\dots,L. \label{sl} 
\eea 
The result follows from the fact that $\l|\varphi\r>$ is an eigenstate of each $\Upsilon^\alpha$,
the $\Upsilon^\alpha$ are self-adjoint, and 
\bea \Upsilon^\alpha U^\beta_p =-U^\beta_p \Upsilon^\alpha ~{~~~~\rm for}~~~~~ 
\alpha\neq \beta. \label{above} \eea  
Note that (\ref{above}) also shows that the Bell subspaces are equivalent.

Let $\P_{mn}$ denote the permutation operator which permutes 
the $m$th and $n$th qubits of the tensor product space $V^L$. These operators 
provide a representation of the symmetric group. Since the $\P_{mn}$ commute with the 
$\Upsilon^\alpha$,  
it follows that each of the subspaces $V^L_{[j:k]}$ is invariant 
under the action of the symmetric group. Thus given any perfect channel, 
it can be used by Alice and Bob to achieve unit fidelity teleportation 
independent of which qubit of the channel is Bob's, how Alice chooses 
to pair the qubits in making the Bell basis measurements, and the 
order in which she performs the measurements. In particular, it is not necessary that 
her first measurement involves the client state. 

We mention here that unlike the $\L=1$ case, the probabilities for the
measurements that may be made by Alice are not necessarily equal in the case of general $\L$. 
An illustration of this fact is given by the example in the Appendix. It is true however that the probability 
a measurement made by Alice is of the Bell class $[j:k]$ is always 1/4, 
independent of $j,\,k$ or $L$. To show this we first construct  
projection operators $P_{[j:k]}$ onto the subspaces $V^L_{[j:k]}$ by 
\bea P_{[j:k]}&=&\frac{1}{4}( I+j\Upsilon^1 )( I+k\Upsilon^2 ) \nn \\
&=& \frac{1}{4}\l(I+ j\Upsilon^1 + k\Upsilon^2 +jk\Upsilon^3 \r). \nn \eea 
Given any perfect channel $|\varphi_{[p:q]}\rangle$ of Bell class $[p:q]$ 
and client state $\l|v\r>$ the density matrix is 
\bea \rho=\l|v\r>\l<v\r| \otimes |\varphi_{[p:q]}\rangle\langle \varphi_{[p:q]}|. \nn \eea
The probability $\P_{[j:k]}$ that Alice makes a measurement of Bell class $[j:k]$ is given by 
\bea \P_{[j:k]}&=& {\rm tr}[(P_{[j:k]}\otimes I)\rho] \nn \\
&=&\frac{1}{4}\l( {\rm tr}[\rho]+{\rm tr}[j(\Upsilon^1\otimes I)\rho]
+ {\rm tr}[k(\Upsilon^2\otimes I)\rho] + {\rm tr}[jk(\Upsilon^3\otimes I)\rho]\r)
\nn \\
&=&\frac{1}{4}\l( 1+{\rm tr}[j(\Upsilon^1\otimes U^1)\rho U^1_L]
+ {\rm tr}[k(\Upsilon^2\otimes U^2)\rho U^2_L] - {\rm tr}[jk(\Upsilon^3\otimes U^3)\rho U^3_L]\r)
\nn \\
&=&\frac{1}{4}\l( 1+{\rm tr}[j(U^1 \otimes \Upsilon^1)\rho U^1_L]
+ {\rm tr}[k(U^2\otimes\Upsilon^2)\rho U^2_L] - {\rm tr}[jk(U^3\otimes\Upsilon^3)\rho U^3_L]\r)
\nn \\
&=&\frac{1}{4}\l( 1+{\rm tr}[jpU_1^1 \rho U^1_L]
+ {\rm tr}[kqU^2_1\rho U^2_L] - {\rm tr}[jkpqU^3_1\rho U^3_L]\r)
\nn \\
&=&\frac{1}{4}\l( 1+jp \l<v|U^1_1|v\r>\langle \varphi_{[p:q]}|U^1_L|\varphi_{[p:q]}\rangle
\r. \nn \\
&&~~~~~~~\l.+ kq \l<v|U^2_1|v\r>\langle \varphi_{[p:q]}|U^2_L|\varphi_{[p:q]}\rangle  
- jkpq \l<v|U^3_1|v\r>\langle \varphi_{[p:q]}|U^3_L|\varphi_{[p:q]}\rangle\r)
\nn \\
&=& \frac{1}{4} \nn \eea 
where the last line follows from (\ref{sl}). 

It is of interest to consider how the above results relate to common physical 
models. It was shown earlier that all singlet states are perfect channels, 
and since they form a subspace (for fixed $L$), they must 
belong to the same Bell class. The ground state of the antiferromagnetic Heisenberg 
model is a singlet, as is the ground state of the one-dimensional Majumdar-Ghosh 
model \cite{mg}, so each is a perfect channel. In one-dimension 
the Heisenberg model is gapless, so any physical realisation would be susceptible 
to errors arising from thermal fluctuations. One way to reduce errors is to 
use a gapped system, which is the case for the Heisenberg model on a 
two-leg ladder lattice \cite{dr} as well as the one-dimensional Majumdar-Ghosh 
model. For the Heisenberg model on the two-dimensional Kagome lattice the 
system is gapless, but the elementary gapless excitations are 
also singlets \cite{ns}. In this instance error due to thermal 
fluctuation is again reduced since all singlet states belong to the 
same Bell class. The existence of such  
subspaces for which all states provide perfect fidelity teleportation 
is reminiscent of decoherence free subspaces 
used to encode logical qubits which are immune to decoherence effects \cite{bklw}.

\subsection{Cluster states}

Cluster states were introduced in \cite{br} as examples of multi-qubit
states with maximal connectedness and high persistency of entanglement. The utilisation of these states for
one-way quantum computation has been studied in
 \cite{rb,man,sch}. We will indicate here how each of the one-dimensional cluster
 states is equivalent to a particular Bell class state. 

The one-dimensional cluster
 states may be 
  defined as the $L$-qubit states
  \bea |\phi^{(L)}\rangle=\frac{1}{2^\L}\l[\bigotimes_{j=1}^{L-1}
  \l(\l|+\r>_j + \l|-\r>_jU^2_{j+1} \r)
  \r]\otimes \l(\l|+\r>_L+\l|-\r>_L\r). \label{cs} \eea
  Consider the set of operators $K_j$ defined by
  \bea
  K_1&=&U^1_1U_2^2, \nn \\
  K_j&=&U^2_{j-1}U^1_j U^2_{j+1}, ~~~~~~j=2,\dots,L-1, \nn \\
  K_L&=&U^2_{L-1}U^1_L . \eea
  It is straightforward to verify these operators satisfy
  \bea [K_j,\,K_l]&=&0~~~~~~~~~~~ j,\,l=1,\dots,L, \nn \\
  K_j \,|\phi^{(L)}\rangle&=& |\phi^{(L)}\rangle~~~~~j=1,\dots,L.   \nn \eea
  Define the operators $G_1$ and $G_2$ by
  \bea
  G_1 &=& \prod_{j=1}^{\L/2}(K_{4j-3}K_{4j}),  \nn \\
  G_2 &=& \prod_{j=1}^{\L/2}(K_{4j-2}K_{4j-1}) \nn \eea
  for $\L/2$ even and
  \bea
  G_1 &=& K_{2\L-1}\prod_{j=1}^{(\L-1)/2}(K_{4j-3}K_{4j}),  \nn \\
  G_2 &=& K_{2\L}\prod_{j=1}^{(\L-1)/2}(K_{4j-2}K_{4j-1}) \nn \eea
   for when $\L/2$ is odd. The operators $G_1$ and $G_2$ necessarily commute and
		moreover
		\bea
		G^1 |\phi^{(L)}\rangle&=&  |\phi^{(L)}\rangle, \nn \\
		G^2 |\phi^{(L)}\rangle&=&  |\phi^{(L)}\rangle. \nn \eea
It can be shown that $G^1,\,G^2$ are equivalent to $\Upsilon^1,\,\Upsilon^2.$ 
We will not give a detailed proof, but rather illustrate some examples.
When $L=6$ we have
   \bea
   G^1&=& U^1_1U^2_2U_3^2U^1_4U^2_5U_4^2U^1_5U^2_6 \nn \\
      &=&  -U^1_1U_2^2U_3^2U_4^3U_5^3U_6^2, \nn \\
      G^2&=& U_1^2U_2^1U_3^2U^2_2U_3^1U_4^2U_5^2U_6^1 \nn \\
	 &=& -U_1^2U_2^3U_3^3U_4^2U_5^2U_6^1
	  \nn \eea
	  whilst in the case $L=8$ we have
	  \bea
	  G^1&=& U^1_1U^2_2U_3^2U^1_4U^2_5U_4^2U^1_5U^2_6U_7^2U_8^1 \nn
	  \\
	     &=&  U^1_1U^2_2U_3^2U^3_4U^3_5U^2_6U_7^2U_8^1 , \nn \\
	     G^2&=&
	     U_1^2U_2^1U_3^2U^2_2U_3^1U_4^2U_5^2U_6^1U_7^2U_6^2U_7^1U_8^2
	     \nn \\
		&=&  U^2_1U^3_2U^3_3U^2_4U^2_5U_6^3U_7^3U_8^2  .    \nn
		\eea
For these instances the equivalence of $G^1,\,G^2$ to $\Upsilon^1,\,\Upsilon^2$ 
can be deduced by inspection. The result holds true not only for all 
linear cluster states, but can be generalised to cluster states defined 
on arbitrary $d$-dimensional lattices as defined in \cite{br}. However, 
the proof is made tedious by the fact that the definition of the $K_j$ 
depends on the choice of cluster in each case, so we omit any details.  		

The fact that each cluster state is equivalent to some perfect channel 
means that it has exactly the same entanglement properties as that channel. 
However, teleportation under our protocol using a cluster state will 
generally fail because the choice of measurement basis is not optimal. 
Mathematically, this is because cluster states do not belong to the stabiliser 
space of $\{\Upsilon^1,\, \Upsilon^2\}$. 
This serves to remind that while entanglement is necessary to 
achieve perfect fidelity teleportation, it is just as necessary 
that the entanglement be {\it ordered} with respect to a choice 
of measurement basis.  In the next section we will construct a 
teleportation-order parameter 
which quantifies this order, and in turn the 
efficiency of a channel. 
 
\section{Teleportation-order}

The manner in which we will construct a teleportation-order 
parameter is motivated by works studying the AKLT model 
and the role of the string-order parameter \cite{pvmc,vmc}. The starting 
point for this study is the concept of localisable entanglement.    

\subsection{Localisable entanglement} \label{le} 

Localisable entanglement is defined as the maximal possible 
entanglement that can be localised between two qubits (or more 
generally qudits), by an optimal choice of measurements on all 
other qubits of the system. The concept of localisable 
entanglement we follow is somewhat 
looser than that of 
\cite{pvmc} in that we do not impose that the measurements 
are local, but rather are Bell measurements.  
Here we will show that each of the basis states
$\l|\r.\vec{j}:\vec{k}\l.\r\}$
has maximal localisable entanglement between {\it any} two qubits 
with respect to {\it any} choice of Bell state measurements
on all the other qubits of the system. Once we have established this
fact, we then show that the same result holds for all states
within a Bell subspace.      

Below,
$$ \l|\r.\vec{j}:\vec{k}\l.\r\}' $$
is defined to be such that
$$\l|\r.\vec{j}:\vec{k}\l.\r\}=
\l|j_1:k_1\r\}\otimes \l|\r.\vec{j}:\vec{k}\l.\r\}'. $$
Now if $\l|\r.\vec{j}:\vec{k}\l.\r\}$ belongs to the Bell class
$[j:k]$ then $\l|\r.\vec{j}:\vec{k}\l.\r\}'$ belongs to the Bell
class $[(jj_1):(kk_1)]$.  Next we appeal to 
(\ref{compact}), which permits us to write
\bea
\l|\r.\vec{j}:\vec{k}\l.\r\}&=&
\l|j_1:k_1\r\} \otimes \l|\r.\vec{j}:\vec{k}\l.\r\}'  \nn \\
&=&\frac{1}{\sqrt{2}}\l(
\l|+,k_1\r>+j_1\l|-,\o{k_1}\r>\r) \otimes \l|\r.\vec{j}:\vec{k}\l.\r\}'
\nn \\
&=& \frac{\kappa}{2^{\L-1/2}}\l|+\r>\otimes \sum_{\vec{p},\vec{q}}
\l|\r.\vec{p}:\vec{q}\l.\r\}\otimes  X^{(jj_1)(kk_1)}_{pq}\l|k_1\r> \nn
\\
&&~~~~~+\frac{j_1\kappa}{2^{\L-1/2}}\l|-\r>\otimes \sum_{\vec{p},\vec{q}}
\l|\r.\vec{p}:\vec{q}\l.\r\}\otimes  X^{(jj_1)(kk_1)}_{pq}\l|\o{k_1}\r>
\nn
\eea
Making pairwise measurements on the interior qubits then projects out a
state
$$
\frac{1}{\sqrt{2}}\l(\l|+\r>\otimes
\l|\r.\vec{p}:\vec{q}\l.\r\}\otimes  X^{(jj_1)(kk_1)}_{pq}\l|k_1\r>
+j_1\l|-\r>\otimes
\l|\r.\vec{p}:\vec{q}\l.\r\}\otimes  X^{(jj_1)(kk_1)}_{pq}\l|\o{k_1}\r>
\r)\nn
$$
It is clear that the two end qubits are disentangled from the rest of
the system by this process, and together form the state
$$(I \otimes X^{(jj_1)(kk_1)}_{pq}) \l|j_1:k_1\r\}$$
which is one of the Bell states.
Using the properties (\ref{p1},\ref{p5})
we may rewrite this  as
\bea (I \otimes X^{(jj_1)(kk_1)}_{pq}) \l|j_1:k_1\r\}
&=&\kappa' (I \otimes X^{jk}_{pq} X^{j_1k_1}_{++})\l|j_1:k_1\r\}
\nn \\
&=&\kappa'' (I \otimes X^{jk}_{pq} X_{j_1k_1}^{++})\l|j_1:k_1\r\}
\nn \\
&=&\kappa'' (I \otimes X^{jk}_{pq}) \l|+:+\r\}
\nn
\eea
This final expression only depends on the Bell class of the channel and
the Bell class of the measurement, so it extends to linear combinations 
of states from the same Bell class. It then  
follows that for any state from a fixed Bell class, keeping in mind 
the the subspace associated with each Bell class is invariant under 
the symmetric group, any 
sequence of Bell measurements on $L-2$ qubits will leave the 
remaining two qubits in a Bell state.  
Depending on the context, it can be said that each Bell class 
state has maximal {\it localisable entanglement} under Bell 
measurements, or maximal {\it entanglement length}.
These concepts have been discussed in \cite{pvmc,vmc} in relation 
to the spin-1 AKLT model with spin-1/2 boundary sites. A significant
feature of this model 
is that the system is gapped with finite-range spin 
correlations, yet has maximal entanglement length.  

Following the notational conventions of \cite{fkr}, the Hamiltonian for the 
AKLT model with spin-1/2 boundaries reads 
\bea H=  h_{1,2} + h_{\L+1,\L}+ \sum_{j=2}^{\L-1} H_{j,j+1}   
\label{aklt} \eea  
where 
\bea h_{j,k}&=&\frac{2}{3}\l(I+\vec{s}_j.\vec{S}_k\r)      \nn \\ 
H_{j,k}&=& \vec{S}_j.\vec{S}_k + \frac{1}{3} \l(\vec{S}_j.\vec{S}_k \r)^2.   
\nn \eea 
Above, $\vec{S}$ is the vector spin-1 operator and $\vec{s}$ 
is the vector spin-1/2 operator. 
The ground state for the system can be constructed exactly using $2\L-2$ virtual qubits
to represent the $\L-1$ local spin-1 spaces
\cite{vmc,fkr}.  
Let 
\bea P^t=P^{++}+P^{-+}+P^{+-}\in {\rm End}(V\otimes V)\nn \eea 
denote the projection onto the triplet space contained in 
$V\otimes V$, and let $P^t_{r}$ denote this operator acting 
on the $r$th and $(r+1)$th qubits of $V^L$. The Hilbert 
space of states for (\ref{aklt}) is the image $W\subset V^L$ of the operator
\bea {\mathbb P}= P_{2}^tP_{4}^t....P^t_{L-2},   \eea  
and the ground state is given by 
\bea\l|AKLT\r>= {\mathbb P}\l|\vec{p}:\vec{q}\r\} \nn 
\eea
where $p_j=q_k=-1\,~\forall j,\,k=1,....,L$; i.e. the ground 
state is the projection of a product of virtual 2-qubit singlet states into $W$.
The ground state is a perfect channel under a protocol which employs a Bell 
measurement on the client state and one boundary spin, followed by a 
sequence of local measurements on the spin-1 sites \cite{pvmc,vmc,fkr}. In 
this procedure the client state is teleported to the other boundary site.
The fact that this protocol works with perfect fidelity can be understood through 
the string-order parameter \cite{pvmc,vmc}.  

For any state $\l|\vartheta\r>\in W$ the string-order parameter $\sop(\vartheta)$ 
is defined as
\bea \sop(\vartheta) =4\left<\vartheta\r|s^z_1 \otimes [\otimes_{k=2}^{\L} \exp(i\pi S^z_k)]
\otimes s_{\L+1}^z\l|\vartheta\r> \label{string} \eea  
which takes values between $-1$ and 1. It can be checked that for the ground state 
\bea \sop(AKLT)=-1. \nn \eea 
We may extend the domain 
of the  local operators $\vec{S}_k$ to act on the direct sum of the triplet and
singlet spaces, and represent each local spin-($1\oplus 0$) space by the full tensor 
product $V\otimes V$ of two virtual qubits.
It is then found that 
$$\exp(i\pi S^z_k)=-U^2 \otimes U^2. $$ 
Therefore the expectation value (\ref{string}) is precisely the
expectation value of $\Upsilon^2$ up to a phase factor of $-(-1)^{\L}$. We thus see that the
expectation value of $\Upsilon^2$ restricted to states in $W$ is equivalent to the string-order parameter
for the case of the AKLT model.

\subsection{Teleportation-order parameter}

By analogy with the string-order parameter, for any state $\l|\Psi\r>\in\,V^L$ we define the 
teleportation-order parameter $\vec{\top}\in {\mathbb R}^3$ to be    
\bea \vec{\top}(\Psi)=\frac{1}{\sqrt{3}}\sum_{j=1}^3 \l<\Psi\r|\Upsilon^j\l|\Psi\r> \vec{e}_j \nn \eea 
where the $\{\vec{e}_j:j=1,2,3\}$ denotes a set of orthonormal 
vectors for ${\mathbb R}^3$. 
Given an arbitrary $\l|\Psi\r>\in\,V^L$ we can make the decomposition 
into a linear combination of representatives from each 
Bell subspace: 
$$\l|\Psi\r>=\sum_{j,k} c_{[j:k]}\l|\Psi_{[j:k]}\r> $$ 
where $\l|\Psi_{[j:k]}\r>\in\, V^L_{[j:k]}$ is assumed to be normalised so that 
$\sum_{j,k} |c_{[j:k]}|^2 =1$.  
We can then determine that 
\bea c_{[j:k]}\l|\Psi_{[j:k]}\r>
&=& P_{[j:k]} \l|\Psi\r> \nn \\
&=&\frac{1}{4}\l(\l|\Psi\r>+ j \Upsilon^1 \l|\Psi\r>
+k\Upsilon^2 \l|\Psi\r> + jk \Upsilon^3 \l|\Psi\r>\r) \label{om} \eea 
which in turn gives 
\bea |c_{[j:k]}|^2=\frac{1}{4}\l( 1+\Omega_{[j:k]}\r) \label{c} \eea 
where we have defined 
\bea \Omega_{[j:k]} = j \l<\Psi\r|\Upsilon^1\l|\Psi\r> 
+k\l<\Psi\r|\Upsilon^2\l|\Psi\r>+jk\l<\Psi\r|\Upsilon^3\l|\Psi\r>.   \label{o} \eea   
Inverting these relations yields 
\bea \l<\Psi\r|\Upsilon^1\l|\Psi\r>&=&\sum_{j,k} j|c_{[j:k]}|^2 \nn \\
\l<\Psi\r|\Upsilon^2\l|\Psi\r>&=&\sum_{j,k} k|c_{[j:k]}|^2 \nn \\
\l<\Psi\r|\Upsilon^3\l|\Psi\r>&=&\sum_{j,k} jk|c_{[j:k]}|^2.   \nn \eea

For any channel we define the {\em efficiency of teleportation} 
$\nabla$ through  
\bea\nabla(\Psi)&=& |\vec{\top}(\Psi)|^2 \nn \\
&=&\frac{1}{3}\sum_{j=1}^3 \l<\Psi|\Upsilon^j|\Psi\r>^2 \nn \\
&=& \frac{1}{3}\left(4\sum_{j,k}|c_{[j:k]}|^4-1\right). \nn \eea   
Since 
\bea (\Upsilon^j)^2= I^{\otimes L} \nn \eea 
then for any state $\l|\Psi\r>$ 
\bea -1\leq \l<\Psi\r|\Upsilon^j\l|\Psi\r> \leq 1, ~~~~~~\forall j=1,\dots,L  \eea 
and so the efficiency takes values in the range $0\leq \nabla  \leq 1$. When $\nabla=0$ 
we see from (\ref{c},\ref{o}) that the state is an equally weighted
linear combination of states from each Bell subspace. At the other
extreme, when $\nabla=1$ it indicates that the state belongs to a 
Bell subspace.

Next we will show that for any product state the efficiency is
bounded: 
$$0\leq \nabla \leq 1/3. $$ 
Let $\l|v_j\r>\in\, V$ be arbitrary and let 
$$\l|w\r>=\l|w_1\r>\otimes \l|w_2\r> \otimes \dots \otimes \l|w_\L\r>. $$  
where 
$$\l|w_j\r>= \l|v_{(2j-1)}\r> \otimes \l|v_{2j}\r>. $$   
It is an exercise to show that 
\bea \l<v_j|U^1|v_j\r>^2 + \l<v_j|U^2|v_j\r>^2 -
\l<v_j|U^3|v_j\r>^2 =1 ~~~~~~~~\forall \,j=1,\dots,L \label{spin}\eea  
and that 
\bea &&1\geq \l<v_j|U^1|v_j\r>^2\geq 0, \nn \\ 
&& 1\geq \l<v_j|U^2|v_j\r>^2 \geq 0, \nn \\
&& 1\geq -\l<v_j|U^3|v_j\r>^2 \geq 0. \nn\eea  
We can then deduce that 
\bea && \l<w_j|U^1\otimes U^1|w_j\r>^2 
+ \l<w_j|U^2\otimes U^2|w_j\r>^2+ 
\l<w_j|U^3\otimes U^3|w_j\r>^2 \nn  \\  
&&~~=\l<v_{(2j-1)}|U^1|v_{(2j-1)}\r>^2 \l<v_{2j}|U^1|v_{2j}\r>^2 
+\l<v_{(2j-1)}|U^2|v_{(2j-1)}\r>^2  \l<v_{2j}|U^2|v_{2j}\r>^2 \nn \\
&&~~~~~~~~~~~~~~~
+ \l<v_{(2j-1)}|U^3|v_{(2j-1)}\r>^2  \l<v_{2j}|U^3|v_{2j}\r>^2 
\nn \\ 
&&~~\leq \l<v_{2j}|U^1|v_{2j}\r>^2+ \l<v_{2j}|U_{2j}^2|v_{2j}\r>^2 
-\l<v_{2j}|U^3|v_{2j}\r>^2 =1 \nn \eea   
and moreover 
\bea 
&&1\geq \l<w_j|U^1\otimes U^1|w_j\r>^2 \geq 0, \nn \\
&&1 \geq \l<w_j|U^2\otimes U^2|w_j\r>^2\geq 0, \nn \\
&&1\geq \l<w_j|U^3\otimes U^3|w_j\r>^2 \geq 0.  \nn  \eea  
Proceeding analogously it follows by an inductive argument that 
$$\sum^3_{j=1} \l<w|\Upsilon^j|w\r>^2 \leq 1 $$ 
and thus $\nabla$ is bounded above by 1/3 for product states. 
Alternatively, if $\nabla>1/3$ for a state $\l|\Psi\r>$ then it must 
certainly be entangled, so $\nabla$ provides a generalised notion  
of an {\it entanglement witness} \cite{hhh,bruss,terhal}.

\subsection{Fidelity}

The above analysis identified channels for which teleportation is achieved with
perfect fidelity; viz. those channels which lie in a Bell subspace. In practise, there may be some error in the channel
which leads to a loss of fidelity. Below we
discuss how such a loss of fidelity may be quantified.

Without loss of generality, since the Bell subspaces are equivalent, let us assume 
that Alice and Bob believe the channel to lie in $V^L_{[+:+]}$. 
The protocol requires that Alice makes
$\L$ Bell measurements on her subsystem which is comprised of the client qubit and $L-1$ qubits of the channel. In the case of perfect channels, 
we have shown that the protocol is independent of the way in which 
she pairs the qubits, nor the 
order in which she makes the measurements. This is also true for the case of non-perfect channels,
which can be seen from eq. (\ref{needed}). So we may simplify the problem by assuming that Alice first makes $\L-1$ Bell measurements on qubits which are contained within the channel, and then the final measurement involving one channel qubit and the client qubit. In view of our earlier discussion on localisable entanglement, the first $\L-1$ measurements project each of the component states  
$\l|\Psi_{[j:k]}\r>$ onto a product of Bell states tensored with a Bell state 
shared by Alice and Bob. Suppose that the result of Alice's first $\L-1$ Bell measurements is of class $[r:s]$.  The Bell class of each of the component Bell states of the total state shared between Alice and Bob after this measurement can be determined from the Bell class of the measurement, as in Sect. \ref{le}. Thus, after Alice's first $\L-1$ Bell measurements, the channel shared by Alice and Bob is found to be of the form 
\bea  \l|\Theta \r> \otimes \l|\Psi \r> \nn \eea 
where $\l|\Theta \r>$ is a state of Bell class $[r:s]$,
onto which Alice has projected as a result of her measurement, 
and
\bea  \l|\Psi \r>= \sum_{j,k} C_{[j:k]} \l|j:k\r\} \nn \eea
with 
\bea C_{[j:k]}= e^{i\theta_{[j:k]}} c_{[rj:sk]}.   \nn \eea 
The above phase factors $\theta_{[j:k]}$ are unknown,  
because Alice's measurement results do not determine the overall phases of 
the components of the remaining shared state. Now the problem has 
been reduced to the investiagation of teleportation across the 2-qubit channel
$\l|\Psi\r>$ which Alice, as a result of her measurements, 
believes to be the Bell state $\l|r:s\r\}$. 

{}From (\ref{comm}) we may write 
\bea \l|v\r>\otimes \l|\Psi\r>=\frac{1}{2}\sum_{p,q}\l|p:q\r\} \otimes
\X_{pq}\l|v\r> \label{imperfect} \eea   
where 
\bea \X_{pq}=\sum_{j,k} C_{[j:k]} \X^{jk}_{pq}. \nn \eea  
Note that $\X_{pq}$ is not necessarily a unitary matrix. The probability $\P_{[pr:qs]}$
of Alice's final measurement result being $(p:q)$, thus making her overall measurement of class
$[pr:qs]$,  is given by 
\bea \P_{[pr:qs]}&=&\frac{1}{4}\l|\l<v\r|\X^\dagger_{pq} \X_{pq}\l|v\r>\r|. 
\nn \eea

Now suppose that Alice's final measurement result is $(p:q)$, so she 
communicates to Bob that the total measurement class is $[pr:qs]$. 
Upon receiving this information,
he would apply the correction gate $(\X_{(pr)(qs)}^{++})^\dagger$ in
attempting to recover the client state. We define the {\it fidelity} 
$\F^{++}_{(pr)(qs)}$ of
this attempted teleportation as the square of the magnitude of the overlap
between the client state and the state Bob has obtained; i.e  
\bea 
\F^{++}_{(pr)(qs)}&=&\frac{\l|\l<v\r|(\X^{(pr)(qs)}_{++})^\dagger \X_{pq}\l|v\r>\r|^2} 
{\l|\l<v\r|\X^\dagger_{pq} \X_{pq}\l|v\r>\r|}  \nn \\
&=&\frac{\l|\l<v\r|(\X^{rs}_{pq})^\dagger \X_{pq}\l|v\r>\r|^2} 
{\l|\l<v\r|\X^\dagger_{pq} \X_{pq}\l|v\r>\r|}  \label{fidelity} \eea 
where we have used (\ref{p1},\ref{p5}). 

We return to eq. (\ref{imperfect}). 
Alice believes that $\l|\Psi\r>$ is
the Bell state $\l|r:s\r\}$ (up to an overall phase which we hereafter
ignore) so we write for real $\theta$  
\bea \l|\Psi\r>&=&(I\otimes R(\theta,\hat{n}))\l|r:s\r\} \nn \eea 
with  
\bea R(\theta,\hat{n})&=&\cos(\theta/2) I -i \sin(\theta/2)
(n_1U^1+n_2U^2+in_3U^3)  \nn \eea 
so that 
\bea \X_{pq}= R(\theta,\hat{n}) \X^{rs}_{pq}. \label{r} \eea 
Using (\ref{x=u}) we then find 
\bea \l|\Psi\r>&=&\l(\cos(\theta/2)I-i \sin(\theta/2)
\l(n_1 X^{r\o{s}}_{rs}+(-1)^s n_2 X^{\o{r}s}_{rs}+(-1)^s
in_3X^{\o{r}\o{s}}_{rs}\r)\r)  
\l|r:s\r\} \nn \eea
such that we can idenitfy  
\bea 
C_{[r:s]}&=& \cos(\theta/2) \nn \\
C_{[r:\o{s}]}&=& -in_1 \sin(\theta/2) \nn \\
C_{[\o{r}:s]}&=& -(-1)^sin_2 \sin(\theta/2) \nn \\
C_{[\o{r}:\o{s}]}&=& (-1)^s n_3 \sin(\theta/2).  \nn \eea   
Normalisation of $\l|\Psi\r>$ requires that $\hat{n}=(n_1,\,n_2,\,n_3)$ is a
unit complex vector. In the case that $\hat{n}$ is real then 
$R(\theta,\hat{n})$ is a unitary matrix corresponding to a rotation of
the Bloch sphere by an angle $\theta$ about an axis determined by
$\hat{n}$. However $R(\theta,\hat{n})$  is not unitary 
for a generic complex unit vector $\hat{n}$.    
Substituting (\ref{r}) into (\ref{fidelity}) gives
\bea
\F^{++}_{(pr)(qs)}
&=&\frac{\l|\l<v\r|(\X^{rs}_{pq})^\dagger R(\theta,\hat{n}) 
\X^{rs}_{pq}\l|v\r>\r|^2}
{\l|\l<v\r|(\X_{pq}^{rs})^\dagger R^\dagger (\theta,\hat{n})R(\theta,\hat{n})
\X^{rs}_{pq}\l|v\r>\r|}.  \nn \eea
The minimum fidelity is 
\bea
{\min} \l(\F^{++}_{(pr)(qs)}\r)
&=& \min_{\l|v\r>}\, \frac{\l|\l<v\r|(\X^{rs}_{pq})^\dagger R(\theta,\hat{n}) 
\X^{rs}_{pq}\l|v\r>\r|^2}
{\l|\l<v\r|(\X_{pq}^{rs})^\dagger R^\dagger (\theta,\hat{n})R(\theta,\hat{n})
\X^{rs}_{pq}\l|v\r>\r|} \nn \\  
&=& \min_{\l|v\r>}\, \frac{\l|\l<v\r| R(\theta,\hat{n}) 
\l|v\r>\r|^2}
{\l|\l<v\r| R^\dagger (\theta,\hat{n})R(\theta,\hat{n})
\l|v\r>\r|} \nn \\
&\geq& 2\cos^2(\theta/2)-1   \label{berry} \eea
where the above inequality holds for all $\hat{n}$. The proof of this result 
is given in Appendix B. We now have   
\bea \min\l(\F^{++}_{(pr)(qs)}\r) &\geq& 2\cos^2(\theta/2)-1 \nn \\
&=& 2 |C_{[r:s]}|^2 -1 \nn \\ 
&=&2 |c_{[+:+]}|^2-1  \nn \\
&=& \frac{1}{2}(\Omega_{[+:+]}-1),  \nn \eea 
so that generally 
\bea \F^{jk}_{pq} &\geq&\frac{1}{2}(\Omega_{[j:k]}-1) .  \nn \eea
That is the quantitity $\Omega_{[j:k]}$, which is simply a linear
combination of the components of the teleportation-order parameter, 
provides a lower bound on the fidelity. The results of simulations for
four-qubit channels are shown in Fig. 2 (also in \cite{bmlm}).  
 
\begin{figure}[h]
\begin{center}
\includegraphics[width=0.8\textwidth]{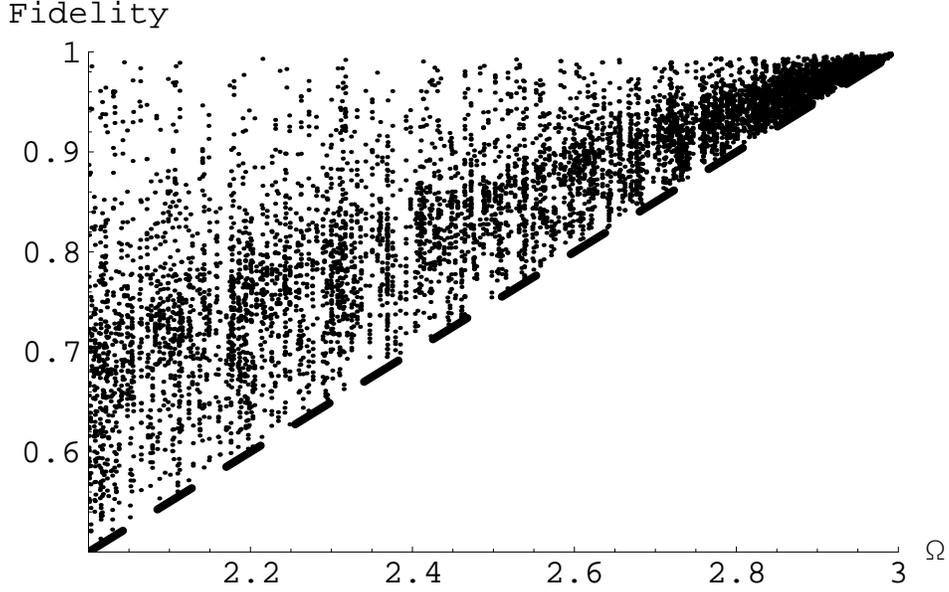}
\end{center}
\caption{Teleportation fidelity versus the quantity $\Omega_{[j:k]}$ 
as defined by (\ref{o}) for all
choices of $[j:k]$. The results shown arise from 2000 simulations, 
with randomly generated client states and 4-qubit channels. The dashed
line denotes
$\F^{jk}_{pq}=(\Omega_{[j:k]}-1)/2$, showing there is a lower bound on
the fidelity 
in terms of the components of the teleportation-order parameter, 
which is independent of the Bell class $[p:q]$ 
of the measurement outcome.  }
\label{fig1}
\end{figure}

\section{Teleportation via three-qubit channels} 

In the above, teleportation was only investigated for an even number of channel qubits. It leaves open the problem of devising a teleportation protocol when the number of channel qubits is odd. Here we won't address this in a general context, but we will show that a protocol does exist for teleportation via three-qubit channels.

We begin by  
defining an orthonormal basis for three-qubit states which generalises the Bell basis. Let
\bea 
\l|+:+:+\r\}&=&\frac{1}{\sqrt{2}}\l(\l|+,+,+\r>+\l|-,-,-\r>\r) \nn \\  
\l|+:+:-\r\}&=&\frac{1}{\sqrt{2}}\l(\l|+,+,-\r>+\l|-,-,+\r>\r) \nn \\
\l|+:-:+\r\}&=&\frac{1}{\sqrt{2}}\l(\l|+,-,+\r>+\l|-,+,-\r>\r) \nn \\
\l|+:-:-\r\}&=&\frac{1}{\sqrt{2}}\l(\l|+,-,-\r>+\l|-,+,+\r>\r) \nn \\
\l|-:+:+\r\}&=&\frac{1}{\sqrt{2}}\l(\l|+,+,+\r>-\l|-,-,-\r>\r) \nn \\
\l|-:+:-\r\}&=&\frac{1}{\sqrt{2}}\l(\l|+,+,-\r>-\l|-,-,+\r>\r) \nn \\
\l|-:-:+\r\}&=&\frac{1}{\sqrt{2}}\l(\l|+,-,+\r>-\l|-,+,-\r>\r) \nn \\
\l|-:-:-\r\}&=&\frac{1}{\sqrt{2}}\l(\l|+,-,-\r>-\l|-,+,+\r>\r). \nn \eea  
Any basis state  can be conveniently expressed as 
\bea  
\l|j:k:l\r\}=\frac{1}{\sqrt{2}}\l(\l|+,k,l\r>+j\l|-,\overline{k},\overline{l}\r>\r)
\label{3qubit} \eea  
and satisfies 
\bea 
\Lambda^1 \l|j:k:l\r\} &=& j \l|j:k:l\r\} \nn \\
\Lambda^2 \l|j:k:l\r\} &=& k \l|j:k:l\r\} \nn \\
\Lambda^3 \l|j:k:l\r\} &=& l \l|j:k:l\r\} \nn \eea 
where 
\bea 
\Lambda^1 &=& U^1\otimes U^1 \otimes U^1 \nn \\ 
\Lambda^2 &=& U^2\otimes U^2 \otimes I \nn \\
\Lambda^3 &=& U^2\otimes I \otimes U^2.  \nn \eea  
Above, the $\Lambda^j$ are mutually commuting self-adjoint operators, so their actions represent a 
simultaneous measurement. We call such a measurement a three-qubit Bell measurement where $(j:k:l)$ 
is the measurement outcome. 
Defining
$Y^{jkl}_{pqr}\in\,{\rm End} (V\otimes V)$ by the relation 
\bea \l|j:k:l\r\}=I \otimes Y^{jkl}_{pqr} \l|p:q:r\r\} \nn \eea   
we find that  
\bea Y^{jkl}_{pqr}= Z^k_q \otimes X^{jl}_{pr}    \nn \eea  
where 
\bea Z^j_j= I, ~~~~~Z^j_{{\overline j}}=U^1 \nn \eea 
and the $X^{jl}_{pr}$ are as before.

Now consider for an arbitrary client state $\l|v\r>=(\alpha\l|+\r>+\beta\l|-\r>)/\sqrt{2}$
\bea 
2\l|v\r>\otimes \l|+:+:+\r\} &=&{\sqrt{2}}\l( \alpha \l|+,+,+,+\r>  
+\beta \l|-,+,+,+\r>\r. \nn \\
&&~~~~~~~\l.+\alpha \l |+,-,-,-\r> +\beta \l |-,-,-,-\r>\r)  \nn \\
&=&\l|+:+:+\r\}\otimes( \alpha \l|+\r> +\beta    \l|-\r>) \nn \\
&&~~~ +\l|+:-:-\r\}\otimes ( \beta \l|+\r>   +\alpha\l|-\r>) \nn \\
&&~~~+\l|-:+:+\r\}\otimes ( \alpha \l|+\r>  -\beta  \l|-\r>) \nn \\
&&~~~+\l|-:-:-\r\}\otimes ( -\beta\l|+\r> +\alpha  \l|-\r>) \nn \\
&=&\l|+:+:+\r\}\otimes \l|v\r>  +\l|+:-:-\r\}\otimes U^1\l|v\r>  \nn \\
&&~~~-+\l|-:+:+\r\}\otimes U^2\l|v\r> +\l|-:-:-\r\}\otimes U^3 \l|v\r>.  \nn 
\eea 
By applying the operators $U^j$ to the last qubit of the space
 we deduce the following relations
\bea
2\l|v\r>\otimes \l|+:-:-\r\} 
&=&\l|+:+:-\r\}\otimes U^1 \l|v\r>  +\l|+:-:+\r\}\otimes \l|v\r>  \nn \\
&&~~~+\l|-:+:-\r\}\otimes U^3\l|v\r> -\l|-:-:+\r\}\otimes U^2 \l|v\r>,  \nn 
\eea
\bea
2\l|v\r>\otimes \l|-:+:+\r\} 
&=&\l|+:+:+\r\}\otimes U^2\l|v\r>  -\l|+:-:-\r\}\otimes U^3\l|v\r>  \nn \\
&&~~~+\l|-:+:+\r\}\otimes \l|v\r> +\l|-:-:-\r\}\otimes U^1 \l|v\r>,  \nn 
\eea
\bea
2\l|v\r>\otimes \l|-:-:-\r\} 
&=&-\l|+:+:-\r\}\otimes U^3\l|v\r>  -\l|+:-:+\r\}\otimes U^2\l|v\r>  \nn \\
&&~~~+\l|-:+:-\r\}\otimes U^1\l|v\r> +\l|-:-:+\r\}\otimes  \l|v\r>.  \nn 
\eea
Applying the operator $U^1$ to the second last qubit of the
space in the above four cases gives 
\bea
2\l|v\r>\otimes \l|+:-:+\r\}
&=&\l|+:+:-\r\}\otimes \l|v\r>  +\l|+:-:+\r\}\otimes U^1\l|v\r>  \nn \\
&&~~~+\l|-:+:-\r\}\otimes U^2\l|v\r> +\l|-:-:+\r\}\otimes U^3 \l|v\r>,  \nn
\eea
\bea
2\l|v\r>\otimes \l|+:+:-\r\}
&=&\l|+:+:+\r\}\otimes U^1 \l|v\r>  +\l|+:-:-\r\}\otimes \l|v\r>  \nn \\
&&~~~+\l|-:+:+\r\}\otimes U^3\l|v\r> -\l|-:-:-\r\}\otimes U^2 \l|v\r>,  \nn
\eea
\bea
2\l|v\r>\otimes \l|-:-:+\r\}
&=&\l|+:+:-\r\}\otimes U^2\l|v\r> -\l|+:-:+\r\}\otimes U^3\l|v\r>  \nn \\
&&~~~+\l|-:+:-\r\}\otimes \l|v\r> +\l|-:-:+\r\}\otimes U^1 \l|v\r>,  \nn
\eea
\bea
2\l|v\r>\otimes \l|-:+:-\r\}
&=&-\l|+:+:+\r\}\otimes U^3\l|v\r>  -\l|+:-:-\r\}\otimes U^2\l|v\r>  \nn \\
&&~~~+\l|-:+:+\r\}\otimes U^1\l|v\r> +\l|-:-:-\r\}\otimes  \l|v\r>.  \nn
\eea
The eight relations above can all be expressed in a unified way: 
\bea  
\l|v\r>\otimes \l|j:k:l\r\}=\frac{1}{2}\sum_{p,q} \l|p:q:(kq)\r\}\otimes
\X^{jl}_{pq} \l|v\r>, \label{3qc}  
\eea  
which is a three-qubit channel generalisation of (\ref{comm}). 
By the same argument as in the two-qubit channel case, we conclude that each $\l|j:k:l\r\}$ is a
perfect channel for teleportation. 
After Alice performs a three-qubit Bell measurement which projects the system onto a state
\bea \l|p:q:(kq)\r\}\otimes
\X^{jl}_{pq} \l|v\r>, \nn \eea  
 two bits of classical information
(i.e. $p$ and $q$)  
need to be transmitted to Bob for him to determine the correction gate.   
 
Because the teleporation protocol requires that only two bits of classical information be sent to Bob, that part of the Bell measurement represented by $\Lambda_2$ becomes redundant. Hence, in analogy with the case where the channel is a state of a system with an even number of qubits, will can still define four Bell classes of channels which give rise to the Hilbert space decomposition
\bea V^3=V^3_{[+:+]}\oplus V^3_{[+:-]}\oplus V^3_{[-:+]}\oplus V_{[-:-]} \nn \eea
where the class indices $[j:k]$ are the eigenvalues of the operators $\Lambda_1$ and $\Lambda_3$.
 Indeed, we can take a generalised form of (\ref{3qc}) 
\bea  \l|v\r>\otimes \sum_k \alpha_k \l|j:k:l\r\}=\frac{1}{2}\sum_{p,q,k} \alpha_k \l|p:q:(kq)\r\}\otimes
\X^{jl}_{pq} \l|v\r>, \label{3qc1}  
\eea  
and perform a reduced three-qubit Bell measurement which is represented by $\Lambda_1$ and 
$\Lambda_3$. Above, $\alpha_1,\,\alpha_2$ are arbitrary up to the constraint of normalisation. 
The consequence of the reduced Bell measurement is that it leaves 
the system in a state of the form 
\bea \sum_k \alpha_k \l|p:q:(kq)\r\}\otimes
\X^{jl}_{pq} \l|v\r>, \nn \eea   
and once again teleportation can be achieved with perfect fidelity. What this result tells us is that the state of the third qubit of the system (i.e. the second qubit of the channel) is of no
consequence in this teleportation protocol. In fact, by a suitable choice of $\alpha_1$ and 
$\alpha_2$ the channel factorises into a tensor product of a Bell state for the first and third qubits of the channel and a disentangled qubit state for the second qubit. Thus this protocol for 
teleportation via a three-qubit channel is essentially a two-qubit channel protocol as the third qubit can be made redundant.   

This raises the question of whether the entanglement of a three-qubit channel can be used to achieve more efficient teleportation than a two-qubit channel. Specifically, can two qubits be
teleported via a three-qubit channel? Within the protocol considered here this is not the case.
Let $\l|v\r>$ and  $\l|w\r>$ be arbitrary qubit states. Using (\ref{3qc})
we calculate
\bea 
\left|v\r>\otimes \l|w\r> \otimes \l|j:k:l\r\} 
&=& \left|v\r> \otimes \l(\frac{1}{2}\sum_{p,q} \l|p:q:(kq)\r\}\otimes
\X^{jl}_{pq} \l|w\r>\r) \nn \\
&=&\frac{1}{4}\sum_{p,q,r,s} \l|r:s:(sq)\r\}\otimes \X^{p(kq)}_{rs}\l|v\r>  \otimes 
\X^{jl}_{pq} \l|w\r>  \nn  \eea 
Now make the change of variable $t=sq$:  
\bea 
&&\left|v\r>\otimes \l|w\r> \otimes \l|j:k:l\r\} \nn \\
&&~~~=\frac{1}{4}\sum_{p,r,s,t} \l|r:s:t\r\}\otimes \X^{p(kst)}_{rs}\l|v\r>
\otimes 
\X^{jl}_{p(st)} \l|w\r>  \nn \\
&&~~~=\frac{1}{\sqrt{8}}\sum_{r,s,t}\l|r:s:t\r\} \otimes \sum_p
\frac{1}{\sqrt{2}}\l(\X^{p(kst)}_{rs} \otimes \X^{jl}_{p(st)}\r)\l(\l|v\r>\otimes
\l|w\r>\r) \nn \\ 
&&~~~=\frac{1}{\sqrt{8}}\sum_{r,s,t}\l|r:s:t\r\} \otimes 
  \l(\X^{+(kst)}_{rs} \otimes
\X^{jl}_{+(st)}\r)\Theta \l(\l|v\r>\otimes
\l|w\r>\r) \nn  
\eea  
where $\Theta = \l(I\otimes I +\kappa \X^{-+}_{++}\otimes \X^{++}_{-+}\r)/\sqrt{2} 
=\l(I\otimes I +\kappa U^2\otimes U^2 \r)/{\sqrt{2}} $. Because $\Theta$ is not invertible, 
there is no possibility to effect  two-qubit teleportation in this manner. 

\section{A qudit generalisation}

As discussed in the original work \cite{bbcjpw}, 
it is also possible to teleport qudit states (see also \cite{werner}).
Let $\{\l|l\r>:\,l=0,\dots, d-1\}$ denote a set of orthonormal basis 
states for a qudit.
For $\omega$ a fixed primitive $d$th root of unity, we introduce the 
permutation and phase matrices defined by 
\bea 
P\l|l\r>&=&\l|l+1 \r>, \nn \\
Q\l|l\r>&=&\omega^l\l|l\r> \nn 
\eea 
and set
\bea
R^{kj}=P^k Q^j . \nn 
\eea 
Throughout, the state labels are taken modulo $d$ so for example $\l|d\r>\equiv \l|0\r>$. 
A qudit generalisation of Bell states is given by 
\bea 
\l| j :k \r\} &=& (I\otimes U^{kj})\l|0:0\r\} \label{ditbell} \\
&=&\frac{1}{\sqrt{d}} \sum_{l=0}^{d-1} \omega^{j l} \l|l\r>\otimes\l| l+k \r> \nn 
\eea
where
\bea 
\l|0:0\r\}=\frac{1}{\sqrt{d}}\sum_{l=0}^{d-1}  \l|l\r>\otimes\l| l\r>. \nn 
\eea 
The generalised Bell states provide a basis which allows us to write 
\bea
\l|j\r>\otimes \l|k\r>=\frac{1}{\sqrt{d}}\sum_{l=0}^{d-1} \omega^{-jl}\l|l:k-j\r\}  . \label{change}
\eea
Letting $\l|v\r>=\sum_{j=0}^{d-1}\alpha_j \l|j\r>$ denote an arbitrary qudit state we find by using
(\ref{change})
\bea
\l|v\r>\otimes \l|0:0\r\}=\frac{1}{d}\sum_{p,q=0}^{d-1} \l|p:{q}\r\} \otimes R^{q\overline{p}}\l|v\r> \nn
\eea
where as before $\overline{p}=-p$. From (\ref{ditbell}) it follows that 
\bea
\l|v\r>\otimes \l|j:k\r\}=\frac{1}{d}\sum_{p,q=0}^{d-1} \l|p:{q}\r\} \otimes \X^{jk}_{p{q}}\l|v\r> \label{ditcomm}
\eea
with 
\bea 
\X^{jk}_{pq}= R^{kj} R^{q\overline{p}}. \nn 
\eea 
It is apparent that (\ref{ditcomm}) is a qudit generalisation of (\ref{comm}). 
As the operators $R^{jk}$ generate a group, since $QP=\omega PQ$, so do the $\X^{jk}_{pq}$. It follows
that our analysis for qubit systems generalises to qudit systems, with the main finding being that
for the $L$-qudit case there are $d^2$ Bell subspaces of perfect channels. The Bell
subspaces are equivalent, with each having dimension $d^{L-2}$.

\section{Conclusion}  

To conclude, we discuss some aspects of our results in the context of 4-qubit channels. The 16-dimensional Hilbert space  $V^4$ decomposes into four Bell subspaces $V^4_{[j:k]}$, each of dimension four. Since these subspaces are all equivalent, we can focus on the space $V^4_{[+:+]}$. This space is precisely the space $G_{abcd}$ of \cite{vddv}, the generic equivalence class of 4-qubit states representing the orbits arising from stochastic local operations and classical communication (see also \cite{lt}). It contains the 4-qubit case of the celebrated  Greenberger-Horne-Zeilinger states, for which it has been argued are the only states exhibiting {\it essential} multi-partite entanglement \cite{wz}. It also contains the state of Higuchi and Sudbery \cite{hs}, which has the largest known average 2-qubit bi-partite entanglement in a 4-qubit system (see also \cite{bssb}). This state, together with its complex conjugate state, provides a basis for the space of singlets contained in $V^4$. Further, there are three states in 
$V^4_{[+:+]}$ which are equivalent to the three 4-qubit cluster states, known to have maximal connectedness and high persistency of entanglement \cite{br}. All the above mentioned states, by representing different forms of multi-partite entanglement, are not equivalent in the sense that they are 
not  related by local unitary transformations. They are however all entirely equivalent for the purpose of teleportation under our prescribed protocol, since they all belong to $V^4_{[+:+]}$. This highlights the fact that the 
entanglement needed to implement this protocol is of a specific type, which depends on each qubit being maximally entangled with the rest of the system (maximal local disorder). Other forms of entanglement the channel might possess are irrelevant. We do emphasise though that for the channel to be effective, this entanglement has to be ordered with respect to the prescribed measurement basis, which is quantified by the teleportation-order parameter.   

Lastly we mention that, besides the one described here, 
there are many possible teleportation protocols which generalise the 
original work of \cite{bbcjpw}; e.g., see \cite{kb,l,lmo}. 
It would be useful  in future work to identify  a correspondence 
between any given teleportation protocol, and a 
teleportation-order parameter which signifies when a channel can be 
used to implement the protocol and effect teleportation with full fidelity. 
Furthermore, the possibilities for performing teleportation without 
a shared reference frame, 
following the ideas developed in \cite{brs}, also warrant investigation.  \\
~~\\ 
%\begin{flushleft} 
{\bf Acknowledgements} - This work was supported by the Australian Research Council. 
We thank Steve Bartlett and Michael Nielsen for helpful advice, and we are indebted to 
Dominic Berry for providing the proof in Appendix B of the inequality (\ref{berry}).
% \end{flushleft}

~~\\~~\\
\centerline{{\bf \Large{Appendix A}}}
~~\\
Here we show by example that not all measurement outcomes are equally likely. Let the channel be 
\bea \l|\Psi\r>=\cos(\phi)\l|+:-\r\}\otimes \l|-:+\r\}
+\sin(\phi)\l|-:+\r\}\otimes\l|+:-\r\} \nn \eea  
which is of Bell class $[-:-]$. Using (\ref{p3},\ref{p4},\ref{needed}) we may write
\bea 4\l|v\r>\otimes \l|\Psi\r> &=&
{\cos(\phi)}\sum_{p_1,p_2,q_1,q_2}\l|\r.p_1 p_2 :q_1 q_2
\l.\r\}\otimes \X^{-+}_{p_2 q_2}\X^{+-}_{p_1q_1} \l|v\r> \nn \\
&&~~~+{\sin(\phi)}\sum_{p_1,p_2,q_1,q_2}\l|\r.p_1 p_2
:q_1 q_2 \l.\r\}\otimes \X^{+-}_{p_2 q_2}\X^{-+}_{p_1q_1}
\l|v\r> \nn \\ 
&=&-{\cos(\phi)}\sum_{p_1,p_2,q_1,q_2}
\ve^{p_1q_1}_{++}\ve^{p_2q_2}_{-+}\l|\r.p_1 p_2 :q_1 q_2
\l.\r\}\otimes X^{++}_{p_2 q_2} X^{--}_{p_1q_1} \l|v\r> \nn \\
&&~~~+{\sin(\phi)}\sum_{p_1,p_2,q_1,q_2}
\ve^{p_1q_1}_{++}\ve^{p_2q_2}_{+-}\l|\r.p_1 p_2
:q_1 q_2 \l.\r\}\otimes X^{++}_{p_2 q_2}X^{--}_{p_1q_1}
\l|v\r> \nn \\ 
&=&-(\cos(\phi)-\sin(\phi))\l|++:++\r\}\otimes
U^3\l|v\r> \nn \\
&&+ (\cos(\phi)+\sin(\phi))\l|++:+-\r\}\otimes
U^2\l|v\r> \nn \\
&&+ (\cos(\phi)-\sin(\phi))\l|++:-+\r\}\otimes
U^2\l|v\r> \nn \\
&&- (\cos(\phi)+\sin(\phi))\l|++:--\r\}\otimes
U^3\l|v\r> \nn \\
&&+ (\cos(\phi)+\sin(\phi))\l|+-:++\r\}\otimes
U^1\l|v\r> \nn \\
&&- (\cos(\phi)-\sin(\phi))\l|+-:+-\r\}\otimes
U^0\l|v\r> \nn \\
&&+ (\cos(\phi)+\sin(\phi))\l|+-:-+\r\}\otimes
U^0\l|v\r> \nn \\
&&- (\cos(\phi)-\sin(\phi))\l|+-:--\r\}\otimes
U^1\l|v\r> \nn \\
&&- (\cos(\phi)-\sin(\phi))\l|-+:++\r\}\otimes
U^1\l|v\r> \nn \\
&&+ (\cos(\phi)+\sin(\phi))\l|-+:+-\r\}\otimes
U^0\l|v\r> \nn \\
&&+ (\cos(\phi)-\sin(\phi))\l|-+:-+\r\}\otimes
U^0\l|v\r> \nn \\
&&- (\cos(\phi)+\sin(\phi))\l|-+:--\r\}\otimes
U^1\l|v\r> \nn \\
&&+ (\cos(\phi)+\sin(\phi))\l|--:++\r\}\otimes
U^3\l|v\r> \nn \\
&&- (\cos(\phi)-\sin(\phi))\l|--:+-\r\}\otimes
U^2\l|v\r> \nn \\
&&+(\cos(\phi)+\sin(\phi))\l|--:-+\r\}\otimes
U^2\l|v\r> \nn \\
&&- (\cos(\phi)-\sin(\phi))\l|--:--\r\}\otimes
U^3\l|v\r>. \nn \eea
{}From the above it can be seen that for any measurement projecting onto a state
\bea  \l|p_1p_2:q_1q_2\r\}\otimes U^j\l|v\r> \nn \eea 
the probability is either  
\bea \frac{1}{16}\l(1+ \sin(2\phi)\r) \nn \eea
or
\bea \frac{1}{16}\l(1- \sin(2\phi)\r) \nn \eea 
so not all measurement outcomes are equally likely.
It is also easily checked for this case that the probability a measurement is of the Bell class 
$[j:k]$ is 1/4, independent of $j$ and $k$, which is consistent with our earlier result. 

\centerline{{\bf \Large{Appendix B}}}
~~\\
Here we prove the inequality (\ref{berry}), viz. 
\bea \min_{\l|v\r>}\, \frac{\l|\l<v\r| R(\theta,\hat{n}) 
\l|v\r>\r|^2}
{\l|\l<v\r| R^\dagger (\theta,\hat{n})R(\theta,\hat{n})
\l|v\r>\r|} 
&\geq& 2\cos^2(\theta/2)-1.   
\nn \eea 
Suppose that $\l|w\r>$ is a vector for which the miminum is achieved. We can then perform a unitary transformation
such that $\l|w\r>$ is transformed into the state $\l|+\r>$. Under such a unitary transformation, the operator
$R(\theta,\hat{n})$ is transformed into $R(\theta,\hat{m})$ for some unit complex vector $\hat{m}$. Importantly, 
the variable $\theta$ is the same for both operators. Setting 
\bea 
\Delta (\theta,\hat{m}) = \frac{\l|\l<+\r| R(\theta,\hat{m}) 
\l|+\r>\r|^2}
{\l|\l<+\r| R^\dagger (\theta,\hat{m})R(\theta,\hat{m})
\l|+\r>\r|} \nn 
\eea 
we now need to show that, for all $\hat{m}$,
\bea \Delta (\theta,\hat{m}) 
&\geq& 2\cos^2(\theta/2)-1.   
\nn \eea 
We may express a generic $ R(\theta,\hat{m}) $ as 
\bea 
R(\theta,\hat{m}) = \cos(\theta/2) I + \sin(\theta/2) \begin{pmatrix}a & c \cr  b & -a\end{pmatrix} 
\nn \eea 
where $a,\,b,\,c$ are complex parameters subject to the normalisation constraint 
\bea  
|a|^2+\frac{1}{2}(|b|^2+|c|^2)=1. \nn 
\eea 
Without loss of generality we may impose $0\leq \theta<\pi $. In terms of these parameters we have
\bea 
\Delta(\theta,\hat{m})
=\frac{|\cos(\theta/2)+a \sin(\theta/2)|^2}{|\cos(\theta/2)+a \sin(\theta/2)|^2+|b\sin(\theta/2)|^2}.
\label{ex1} \eea 
For any fixed $a$, (\ref{ex1}) is minimised by maximising $|b|^2$. We thus choose $c=0$ leading to 
\bea 
\Delta(\theta,\hat{m})&=&\frac{|\cos(\theta/2)+a \sin(\theta/2)|^2}{|\cos(\theta/2)+a \sin(\theta/2)|^2+2(1-|a|^2)\sin^2(\theta/2)} \nn \\
&=&\frac{ \cos^2(\theta/2)+|a|^2 \sin^2(\theta/2)+2\Re (a)\sin(\theta/2)\cos(\theta/2) }
{1+(1-|a|^2) \sin^2(\theta/2)+2\Re (a)\sin(\theta/2)\cos(\theta/2)}
\label{ex2} \eea 
where $\Re (a)$ denotes the real part of $a$. The above expression is minimised when $a$ is real and given by 
\bea 
a= \frac{\sqrt{1-4\cos^2(\theta/2)\sin^2(\theta/2)}-1}{2\sin(\theta/2)\cos(\theta/2)}
=\frac{\cos(\theta)-1}{\sin(\theta)}  . \label{ex3}
\eea 
Finally, substituting (\ref{ex3}) into (\ref{ex2}) gives
\bea 
\min_{\hat{m}}\l(\Delta(\theta,\hat{m})\r)=\cos(\theta)=2\cos^2(\theta/2)-1 \nn 
\eea 
as required.

\end{document}